\date{100408}
\documentclass[showpacs,twocolumn,floatfix]{revtex4}
\usepackage{graphicx,psfrag,amsmath,amssymb,amsfonts,bbm,latexsym,color,dcolumn,
epsf,graphpap}

\definecolor{red}{rgb}{1,0,0}
\definecolor{blue}{rgb}{0,0,1}
\definecolor{skyblue}{rgb}{0,0,.5}
\definecolor{green}{rgb}{0,1,0}
\definecolor{orange}{cmyk}{0,.4,1,0}

\begin{document}
\title{Numerical evaluation of the Casimir interaction between cylinders }

\author{F.C. Lombardo \footnote{lombardo@df.uba.ar}}
\author{F.D.  Mazzitelli\footnote{fmazzi@df.uba.ar}}
\author{P.I. Villar \footnote{paula@df.uba.ar}}

 \affiliation{Departamento de F\'\i sica {\it Juan Jos\'e
 Giambiagi}, FCEyN UBA, Facultad de Ciencias Exactas y Naturales,
 Ciudad Universitaria, Pabell\' on I, 1428 Buenos Aires, Argentina}

\date{today}
\begin{abstract}
We numerically evaluate the Casimir interaction energy for
configurations involving two perfectly conducting  eccentric
cylinders and a cylinder in front of a plane. We consider in
detail several special cases. For  quasi-concentric cylinders, we
analyze the convergence of a perturbative evaluation based on
sparse  matrices. For concentric cylinders,  we obtain
analytically  the corrections to the proximity force approximation
up to second order, and we  present an improved numerical
procedure to evaluate the interaction energy at very small
distances. Finally, we consider the configuration of a cylinder in
front of a plane. We first show numerically that, in the
appropriate limit, the Casimir energy for this configuration can
be obtained from that of two eccentric cylinders. Then we compute
the interaction energy at small distances, and compare the
numerical results with the analytic predictions for the first
order corrections to the proximity force approximation.

\end{abstract}


\maketitle

\section{Introduction}
It has been 60 years since Casimir \cite{Casimir} found a profound
explanation for the retarded van der Waals interaction as a
manifestation of the zero-point energy of the quantum
electromagnetic field. For many years the Casimir effect was
little more than a theoretical curiosity. But interest in the
phenomenon has blossomed in recent years. Experimental physicists
have realized that the Casimir
force affects the workings of
micromachined devices, while advances in instrumentation have
enabled the force to be measured with ever-greater accuracy. On
theoretical grounds,  considerable progress has also been achieved
by studying the dependence of the Casimir force with the geometry
of the conducting surfaces \cite{Todos}.

Up to now, most experiments aiming at a  measurement of the
Casimir force have been performed with parallel plates \cite{pp},
or with a sphere in front of a plane \cite{experiments}. The parallel
plates configuration has a stronger signal, but the main
experimental difficulty is to achieve parallelism between the
plates. This problem is of course not present in the case of a
sphere in front of a plane, but its drawback is that the force is
several orders of magnitude smaller. The problem of the theoretical
evaluation of the electromagnetic force for this configuration
has been solved recently \cite{esfera}.

The configuration of two eccentric cylinders have both experimental
and theoretical interest \cite{europhysics,mazzitelli04}.
 Although parallelism is as difficult as for the plane-plane configuration,
the fact that the concentric configuration is an unstable
equilibrium position opens the possibility of measuring the
derivative of the force using null experiments (for example, one
could consider experimental configurations in which a metallic
wire is placed inside a larger hollow cylinder). The Casimir
interaction energy between two eccentric cylindrical shells has
been computed in \cite{NJP}, and was  initially reported in
\cite{PRA74}. Therein, it was used  the mode summation technique
combined with the argument theorem in order to write the Casimir
energy as a contour integral in the complex plane, to end with an
exact formula in which the vacuum energy is written in terms of
the determinant of an infinite matrix. As a bonus, it has also
been shown that the matrix elements in  the general formula  for
two eccentric cylinders reproduce, as a limiting case of
relevance, those of the  Casimir energy for the cylinder-plane
configuration. The latter geometry is also of experimental
interest: being intermediate between the sphere-plane and the
plane-plane geometries,  it can shed some light on the
longstanding controversy about thermal corrections to the Casimir
force. Keeping the two plates parallel has proved very difficult,
while the sphere and plate configuration avoids this problem, the
force is not extensive. In the case of the cylinder-plane
configuration, it is easier to hold the cylinder parallel and the
force results extensive in its length. There is an ongoing
experiment to measure the Casimir force for this configuration
\cite{hayespra72}.

The aim of this paper
is to provide a precise numerical evaluation of the analytical results obtained in \cite{NJP}.
The numerical evaluations will allow as to test  different approximations, as the proximity
force approximation (PFA) for close surfaces, and  the  "around the diagonal"  approximation
for quasi concentric cylinders. We will also show numerically that the energy of the
eccentric cylinders
configuration reproduces that of the cylinder-plane configuration, in the limit of
very large eccentricity, when the radius of the external cylinder is also very large.
Finally, we will present a detailed numerical evaluation for the vacuum energy
for the  cylinder-plane configuration, providing numerical support for the analytic predictions
of the first order corrections
to PFA \cite{Bordag}.

The paper is organized as follows. In Section \ref{numeric} we
will discuss the general procedure used for the numerical
evaluation of the Casimir energy in every case considered. In
Section \ref{eccentric}, we will evaluate the exact formula for
the interaction energy between eccentric cylinders. The complexity
of the numerical evaluation increases  as the radii of the
cylinders get closer, and we provide details of the size of the
matrices needed to assure convergence of the numerical results. In
Section \ref{quasi} we  analyze the particular case of
quasi-concentric cylinders. We test the validity of the
approximation developed in \cite{NJP}, based on tridiagonal
matrices, that we extend to the next order by considering
pentadiagonal matrices. In Section \ref{concentric} we consider
the particular case of concentric cylinders. We will obtain a new
analytic result in the small distance limit, that includes the
corrections to PFA up to the second order. We will also present an
improved numerical method to evaluate the interaction energy at
small distances. Finally, in section \ref{cp}, we will numerically
show that the interaction energy for the cylinder-plane
configuration can be derived in the appropriate limit from the
eccentric cylinders configuration, a result that was anticipated
analytically for the matrix elements in \cite{NJP}. In addition,
we will evaluate numerically the cylinder-plane Casimir energy as
the minimum distance between the surfaces is much smaller than the
radius of the cylinder. We will be able to show numerically that
the energy is well reproduced in this limit by the PFA, and to
compute the first order correction to PFA for both TM and TE
modes. In the first case (TM), the fits of the numerical data
reproduce with high precision the analytic prediction
\cite{Bordag}. On the other hand, the fits for TE modes are close
to  the analytic results or not, depending on the assumption about
the higher order corrections.

\section{Numerical Approach}
\label{numeric}

\begin{figure}
\includegraphics[width=5.5cm]{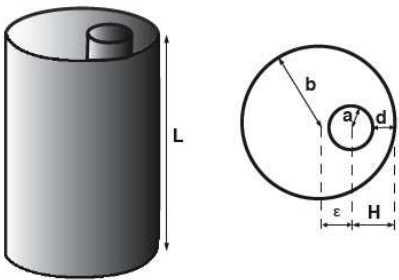}
\caption{Geometrical configuration for the eccentric cylinders. Two perfectly
conducting cylinders of radii $a < b$, length $L \gg a,b$, and eccentricity $\epsilon$
interact via the Casimir force.}
\label{fig1}
\end{figure}

The evaluation of the Casimir interaction energy between two eccentric
cylindrical shells (Fig.\ref{fig1}) has been initially performed using
PFA in Ref.\cite{europhysics}. However, it is possible to go beyond the
PFA and find an exact formula for the interaction energy \cite{NJP,PRA74}.
This can be done using a mode by mode summation technique combined
with the argument theorem.  By starting with the expression of the
Casimir energy as $E=(\hbar/2) \sum_p(\omega_p-\tilde{\omega}_p)$,
it has been shown in  \cite{NJP} that the Casimir interaction energy for two
eccentric cylinders can be written as
\begin{eqnarray}
E_{12} &= & \frac{L}{4 \pi  a^2} \int_0^{\infty} d\beta
~\beta ~{\rm ln}( M(\beta)) \nonumber \\
&=& \frac{L}{4 \pi  a^2} \int_0^{\infty}  d\beta ~\beta
~\bigg[{\rm ln}(M^{\rm TE}(\beta))
+ {\rm ln}(M^{\rm TM}(\beta))]  \nonumber \\
&=& E^{\rm TE}+E^{\rm TM},
\label{E12}
\end{eqnarray}
where $M^{\rm TM}(\beta)={\rm det}[\delta_{n p}- A_{n,p}^{\rm
TM}]$ and $M^{\rm TE}(\beta)={\rm det}[\delta_{n p}- A_{n,p}^{\rm
TE}]$. Here $\beta $ is a dimensionless integration variable and
$n,p$ are arbitrary integers. Roughly speaking, the function $M$
that determines the Casimir energy through Eq.(\ref{E12}) is such
that its zeros give the eigenfrequencies of the geometric
configuration. More precisely, it is  the ratio of the function
associated to the actual geometric configuration and the one
associated to a configuration in which the conducting surfaces are
very far away from each other \cite{NJP}. The matrices
$A_{n,p}^{\rm TM}$ and $A_{n,p}^{\rm TE}$ are defined as
\begin{eqnarray}
A_{n,p}^{\rm TM}=\frac{I_{n}(\beta)}{K_n(\beta)}
\sum_m \frac{K_m(\alpha \beta)}
{I_m(\alpha \beta)} I_{n-m}\bigg(\beta
\frac{\epsilon}{a}\bigg)
I_{p-m}\bigg(\beta \frac{\epsilon}{a}\bigg),
\label{FullTM}
\end{eqnarray}
and similarly, for the $\rm TE$ modes, we have
\begin{eqnarray}
A_{n,p}^{\rm TE}=\frac{I'_{n}(\beta)}{K'_n(\beta)}
\sum_m \frac{K'_m(\alpha \beta)}
{I'_m(\alpha \beta)} I_{n-m}\bigg(\beta
\frac{\epsilon}{a}\bigg)
I_{p-m}\bigg(\beta \frac{\epsilon}{a}\bigg),
\label{FullTE}
\end{eqnarray}
where  $\alpha = b/a$ is the radio between the outer and inner
cylinder's radii and $\epsilon$ is the eccentricity (see Fig.1).
$I_n$ and $K_n$ denote the modified Bessel functions.

In order to calculate the Casimir interaction energy, one needs to
perform a numerical evaluation of the determinants in
Eq.(\ref{E12}), followed by a numerical integration in the
variable $\beta$.  We find that as $\alpha $ approaches small
values, larger matrices are needed for ensuring convergence.
Likewise, as $\alpha \rightarrow 1$, the contribution to the
integral is significative for a bigger integration range (bigger
values of $\beta$ contribute). That turns the problem into a real
challenge from the numerical point of view. We numerically compute
the Casimir interaction energy using a Fortran program. Once the
$M$ matrix elements for each configuration considered is defined,
we use a standard routine to calculate its eigenvalues and
determinant. Finally, we  perform a standard integration over all
values of $\beta$.  The parameters used by the program are: the
dimension of the $M$ matrix (N,N), the number of addends $m$
corresponding to each element of the M matrix, the integration
limit ($\beta_{\rm max}$) and the precision desired. The
difficulty in running the programme lays in the compromise taken
between all the parameters chosen.

In the following, we will evaluate the Casimir interaction energy for
eccentric, quasi concentric, concentric cylinders, and also for the
particular limit of a cylinder in front of a plane.

\section{Eccentric cylinders}
\label{eccentric}

In this section, we present the numerical results for  the Casimir
interaction energy for two eccentric cylinders given by
Eqs.(\ref{FullTM}) and (\ref{FullTE}).

\begin{figure}[!ht]
\centering
\includegraphics[width=8.7cm]{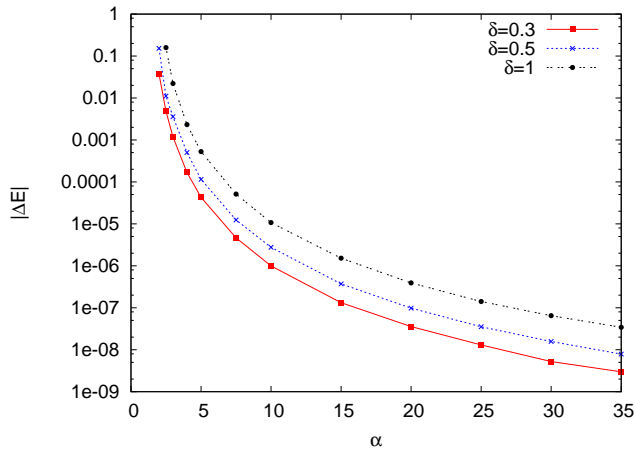}
\caption{Exact Casimir interaction energy difference $|\Delta E|$
between the eccentric and concentric configurations as a function
of $\alpha=b/a$ for different values of $\delta=\epsilon/a$. Here
$\Delta E=E_{12}-E_{12}^{cc}$. Energies are measured in units of
$L/4\pi a^2.$} \label{NJPfig3}
\end{figure}

\begin{figure}[!ht]
\centering
\includegraphics[width=8.7cm]{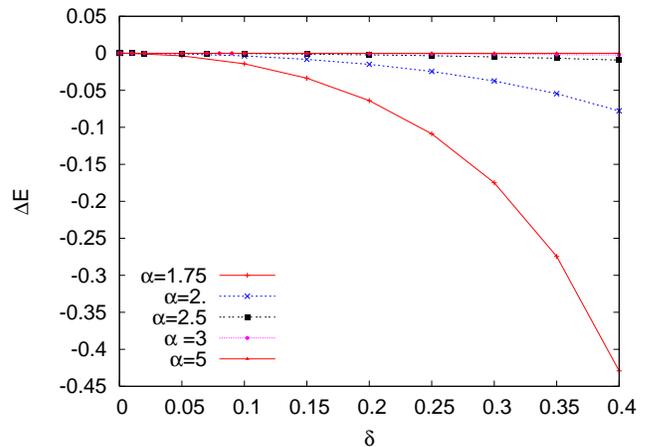}
\caption{Exact Casimir interaction energy difference $\Delta E$
between the eccentric and concentric configurations as a function
of $\delta=\epsilon/a$ for different values of $\alpha=b/a$.
Energies are measured in units of $L/4\pi a^2$. The maximum at
$\delta=0$ shows the instability of the concentric equilibrium
position.} \label{NJPfig4}
\end{figure}

In Figs.\ref{NJPfig3} and \ref{NJPfig4}, we reproduce
the exact Casimir interaction energy difference
 $\Delta E= E_{12}-E_{12}^{cc}$ between
the eccentric and concentric configurations. In Fig.\ref{NJPfig3}
we plot the interaction energy difference $|\Delta E|$ as a
function of $\alpha$ for different values of the eccentricity
$\delta=\epsilon/a$. These numerical results interpolate between
the PFA and the asymptotic behavior for large $\alpha$ \cite{NJP}.
In Fig.\ref{NJPfig4} we show the Casimir interaction energy
difference as a function of $\delta$ for various values of
$\alpha$. Again, it is evident that the equilibrium position
($\delta=0$) is unstable.

In Refs.\cite{NJP,PRA74} similar plots were performed using an
algebraic evaluation of the trace of the matrix $M$, and a
numerical integration, both using Mathematica. Due to this
procedure, it was possible to evaluate the vacuum energy  only for
relatively large values of the parameter $\alpha$,  in order to
reach convergence. With the numerical method we are presenting
here, we are able to include smaller values of $\alpha$, closer to
the PFA region, where previous numerical calculations could not
reach. Both in Fig.\ref{NJPfig3} and Fig.\ref{NJPfig4} we include
runs for $\alpha > 1.75$. In order to achieve so, we have used
matrices of dimension (21,21) and 501 addends in the sums of Eqs.
(\ref{FullTM}) and (\ref{FullTE}) to assure convergence
 (variation smaller than $10^{-4}$). For values $\alpha > 5$,
smaller $M$ matrices (5,5) can be used to obtain the plots with equal
precision (indeed, as it was shown in \cite{NJP}, for $\alpha \rightarrow \infty$
the energy is dominated by the 00-element of the matrix $M$). The size of $M$ for the runs was
set by the smaller values of $\alpha$ that needed bigger $M$ matrix
to obtain the same accuracy.

The convergence of the numerical results depends both on the
values of $\alpha$ and $\delta$. For example, for $\alpha = 3$ and
$\delta = 0.01$ matrices (5,5) are enough, while for $\delta = 1$
matrices (9,9) are needed. In the case of $\alpha = 1.25$, it is
necessary to use matrices of dimensions (55,55) and (101,101) when
$\delta$ is 0.01 and 0.1, respectively.

\section{Quasi-concentric cylinders}
\label{quasi}

In this section, we  consider the situation
in which the eccentricity of the
configuration is much smaller than the radius of the inner cylinder ($\delta \ll 1$).
For  a small non-vanishing eccentricity,
the behaviour of the Bessel functions in Eqs.(\ref{FullTM})
and (\ref{FullTE})
is $I_{m-n}(\beta \delta) \sim (\beta \delta)^{n-m}$
for small arguments. This
suggests that the main contribution should be the one coming from
the diagonal elements, and that one only needs to use matrix elements
near the diagonal. We will test this idea through a  numerical comparison
between the Casimir interaction energy for different
approximations and the exact energy derived in the
previous section.

{\it First order approximation}.
To begin with, we will only consider the matrix elements proportional
to $I_0(\beta\delta), I_1(\beta\delta)$ and $I_1^2(\beta\delta)$ as we are assuming small eccentricity
$\delta=\epsilon /a \ll1$. In this particular case, the $M$ matrix
become tridiagonal and the $\epsilon$-dependent part of the Casimir
energy will be quadratic in the eccentricity.
We will describe in detail the case of the Dirichlet  (TM) modes; the treatment
of Neumann (TE) modes is similar. As was already mentioned in \cite{NJP},
to order ${\cal O}(\delta^2)$, the
non-vanishing elements of the matrix $A_{np}^{\rm TM}$ are:
\begin{eqnarray}
A_{n,n}^{\rm TM(1)} & \simeq & \frac{I_{n}(\beta)}{K_n(\beta)}
\bigg[ \frac{K_n(\alpha \beta)}
{I_n(\alpha \beta)} I_0^2(\delta \beta) +  \frac{K_{n-1}(\alpha \beta)}
{I_{n-1}(\alpha \beta)} I_1^2(\delta \beta) \nonumber \\
&+&  \frac{K_{n+1}(\alpha \beta)}
{I_{n+1}(\alpha \beta)} I_1^2(\delta \beta) \bigg],  \nonumber \\
A_{n,n+1}^{\rm TM(1)} & \simeq & \frac{I_{n}(\beta)}{K_n(\beta)}
\bigg[ \frac{K_n(\alpha \beta)}
{I_n(\alpha \beta)} +  \frac{K_{n+1}(\alpha \beta)}
{I_{n+1}(\alpha \beta)}\bigg] I_0(\delta \beta)  I_1(\delta \beta),
\nonumber \\
A_{n+1,n}^{\rm TM(1)} & \simeq & \frac{I_{n+1}(\beta)}{K_{n+1}(\beta)}
\bigg[ \frac{K_n(\alpha \beta)}
{I_n(\alpha \beta)} +  \frac{K_{n+1}(\alpha \beta)}
{I_{n+1}(\alpha \beta)}\bigg] \nonumber \\
&\times & I_0(\delta \beta)  I_1(\delta \beta).\nonumber
\end{eqnarray}

{\it Second order approximation}.
In this case, we will consider that the main contribution
to the Casimir interaction energy comes from the terms
that contain up to ${\cal O}(\delta^4)$, extending the
previous approximation to the next non trivial order. Then, the  matrix A
has additional non diagonal
contributions, i.e. it is a pentadiagonal matrix with non-vanishing
elements given by
\begin{eqnarray}
A_{n,n}^{\rm TM(2)} & \simeq & \frac{I_{n}(\beta)}{K_n(\beta)}
\bigg[ \frac{K_n(\alpha \beta)}
{I_n(\alpha \beta)} I_0^2(\delta \beta) +  \frac{K_{n-1}(\alpha \beta)}
{I_{n-1}(\alpha \beta)} I_1^2(\delta \beta) \nonumber \\
&+&  \frac{K_{n+1}(\alpha \beta)}
{I_{n+1}(\alpha \beta)} I_1^2(\delta \beta)  + \frac{K_{n-2}(\alpha \beta)}
{I_{n-2}(\alpha \beta)} I_2^2(\delta \beta) \nonumber \\
&+&  \frac{K_{n+2}(\alpha \beta)}
{I_{n+2}(\alpha \beta)} I_2^2(\delta \beta)  \bigg],  \nonumber \\
A_{n,n+1}^{\rm TM(2)} & \simeq & \frac{I_{n}(\beta)}{K_n(\beta)}
\bigg[ \frac{K_n(\alpha \beta)}
{I_n(\alpha \beta)} +  \frac{K_{n+1}(\alpha \beta)}
{I_{n+1}(\alpha \beta)}\bigg] I_0(\delta \beta)  I_1(\delta \beta),\nonumber \\
A_{n+1,n}^{\rm TM(2)} & \simeq & \frac{I_{n+1}(\beta)}{K_{n+1}(\beta)}
\bigg[ \frac{K_n(\alpha \beta)}
{I_n(\alpha \beta)} +  \frac{K_{n+1}(\alpha \beta)}
{I_{n+1}(\alpha \beta)}\bigg]\nonumber \\ &\times & I_0(\delta \beta)  I_1(\delta \beta),\nonumber \\
A_{n,n+2}^{\rm TM(2)} & \simeq & \frac{I_{n}(\beta)}{K_n(\beta)}
 \bigg\{\bigg[ \frac{K_n(\alpha \beta)}
{I_n(\alpha \beta)} +  \frac{K_{n+2}(\alpha \beta)}
{I_{n+2}(\alpha \beta)}\bigg] \nonumber \\ &\times & I_0(\delta \beta)  I_2(\delta \beta)  \nonumber \\
&+& \frac{K_{n+1}(\alpha \beta)}{I_{n+1}(\alpha \beta)}
I_1^2(\delta \beta) \bigg\} , \nonumber \\
A_{n+2,n}^{\rm TM(2)} & \simeq & \frac{I_{n+2}(\beta)}{K_{n+2}(\beta)}
\bigg\{\bigg[ \frac{K_n(\alpha \beta)}
{I_n(\alpha \beta)} +  \frac{K_{n+2}(\alpha \beta)}
{I_{n+2}(\alpha \beta)}\bigg]\nonumber \\ &\times & I_0(\delta \beta)
 I_2(\delta \beta)  \nonumber \\
&+& \frac{K_{n+1}(\alpha \beta)}{I_{n+1}(\alpha \beta)}
 I_1^2(\delta \beta) \bigg\} . \nonumber
\end{eqnarray}

\begin{figure}[!ht]
\centering
\includegraphics[width=8.7cm]{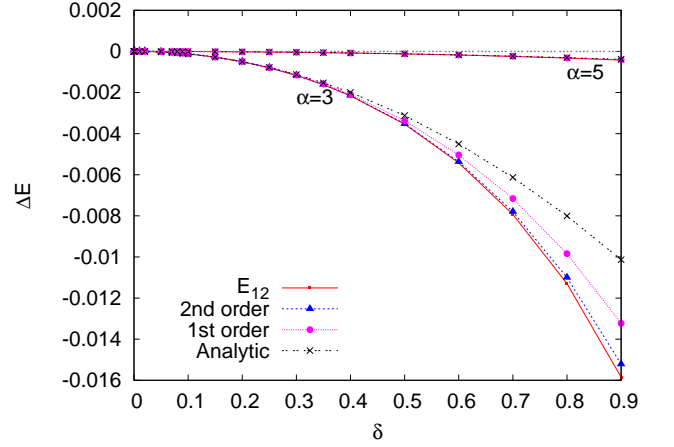}
\caption{Exact Casimir interaction energy difference $\Delta E$
between the eccentric and concentric configurations as a function
of  $\delta=\epsilon/a$ for different values of $\alpha$. The
solid line is the exact evaluation of the Casimir interaction
energy, while the dashed line with triangles is the second order
approximation and the dashed line with dots is the first order
one. The analytic curve (dashed line with crosses) is the result
of using Eq.(\ref{ec48}). Energies are measured in units of
$L/4\pi a^2$.} \label{cuasifig1}
\end{figure}

In the following, we will  numerically
compare the Casimir interaction energy for the quasi
concentric configuration computed using the first  and second
order approximations described above, with the one
obtained using the exact formula given in Eqs.(\ref{FullTM})
and (\ref{FullTE}).

\begin{figure}[!h]
\centering
\includegraphics[width=8.7cm]{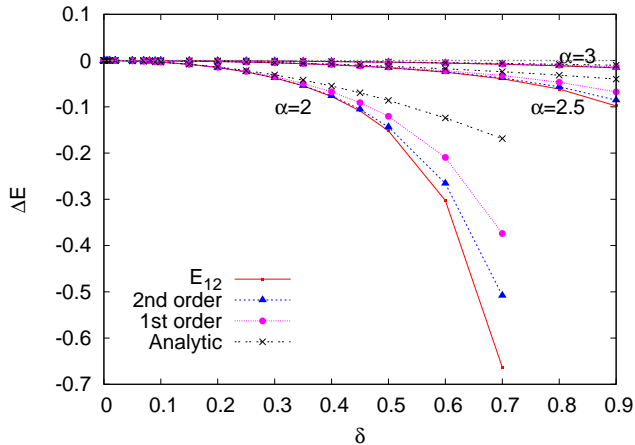}
\caption{Exact Casimir interaction energy difference $\Delta E$
between the eccentric and concentric configurations as a function
of  $\delta=\epsilon/a$ for different values of $\alpha$. The
solid line is the exact evaluation of the Casimir interaction
energy, while the dashed line with triangles is the second order
approximation and the dashed line with dots is the first order
one. The analytic curve is the result of using Eq.(\ref{ec48}).
Energies are measured in units of $L/4\pi a^2$.} \label{cuasifig2}
\end{figure}

\begin{figure}[!ht]
\centering
\includegraphics[width=8.7cm]{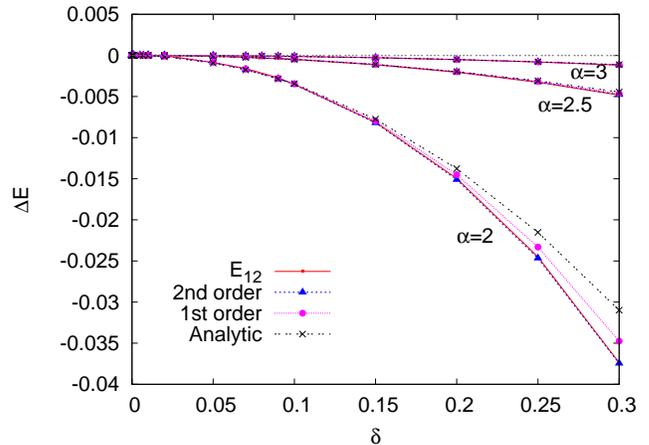}
\caption{Exact Casimir interaction energy difference $\Delta E$
between the eccentric and concentric configurations as a function
of  $\delta=\epsilon/a$ for different values of $\alpha$ (a
smaller range of $\delta$). The solid line is the exact evaluation
of the Casimir interaction energy, while the dashed line with
triangles is the second order approximation and the dashed line
with dots is the first order one. The analytic curve (dashed line
with crosses) is the result of using Eq.(\ref{ec48}). Energies are
measured in units of $L/4\pi a^2$.} \label{cuasifig3}
\end{figure}

In Figs.\ref{cuasifig1}, \ref{cuasifig2} and \ref{cuasifig3} we
present the Casimir interaction energy diference $\Delta
E=E_{12}-E_{12}^{cc}$ between the eccentric and concentric
configurations as a function of $\delta=\epsilon/a$ for different
values of $\alpha$. Therein, we can see the different curves
corresponding to the exact energy difference (solid line) and the
approximated ones obtained to first (dashed line with dots) and
second order (dashed line with triangles) in the eccentricity. In
all cases, matrix of
 dimension (21,21)\cite{nota}
and 501 addends in the sums have been used to assure convergence.

We are also comparing the numerical results with the "analytic"
result for quasi-concentric cylinders obtained Ref.\cite{NJP}.
Therein, it was shown that the determinant of the tridiagonal
matrix can be explicitly evaluated up to quadratic order in
$\delta$, and therefore  it was possible to write the interaction
energy as a series
\begin{eqnarray}
E_{12}^{\rm TM} &=& E_{12}^{{\rm TM, cc}} - \frac{L \epsilon^2}{4
\pi a^4} \sum_n  \int_0^{\infty} d\beta \; \beta^3
\frac{1}{1-{\cal D}^{\rm TM, cc}_{n,n}}
\nonumber \\
&\times&
 \left[ {\cal D}^{\rm TM}_n +
\frac{{\cal N}^{\rm TM}_n}{1-{\cal D}^{\rm TM, cc}_{n+1,n+1}} \right].
\label{ec48}
\end{eqnarray}
Here
\begin{eqnarray}
{\cal D}^{\rm TM}_n & \equiv & \frac{ {\cal D}^{\rm TM, cc}_{n,n}}{2} + \frac{I_n(\beta)}{4 K_n(\beta)} \left[
\frac{K_{n-1}(\alpha\beta)}{I_{n-1}(\alpha \beta)} +
\frac{K_{n+1}(\alpha \beta)}{I_{n+1}(\alpha \beta)} \right] , \nonumber \\
{\cal N}^{\rm TM}_n & \equiv & \frac{I_n(\beta) I_{n+1}(\beta)}{4
K_n(\beta) K_{n+1}(\beta)} \left[ \frac{K_{n}(\alpha
\beta)}{I_{n}(\alpha \beta)} + \frac{K_{n+1}(\alpha
\beta)}{I_{n+1}(\alpha \beta)} \right]^2 \, ,\nonumber\\
{\cal D}_{n,n}^{\rm TM, cc} &=& \frac{I_n(\beta)}{K_n(\beta)}
\frac{K_n(\alpha\beta)}{I_n(\alpha\beta)}\, .
 \label{newd}
\end{eqnarray}
The contribution of the TE modes to the interaction energy
$E_{12}^{\rm TE}$ has a similar expression, replacing the Bessel
functions by their derivatives with respect to the argument in the
equations above. The numerical evaluation of this formulae in the
plot is presented as the ``analytic" curve.

In Fig.\ref{cuasifig1} we can see that the difference between
several approximations increases with $\delta$, for a given value
of $\alpha$. For example, for $\alpha = 5$ and $\delta = 0.5$, the
analytic approximation differs from the exact result in $2\%$, and
the first order approximation in $1\%$. For a larger value of
$\delta$, as $\delta = 0.9$, the differences are $9\%$ and $4\%$
for the analytic and first order approximations, respectively. The
results for the second order approximation coincide with the exact
result within the accuracy imposed. However, as the value of
$\alpha$ becomes smaller, the differences become more visible. For
$\alpha = 3$ the accuracy of the different approximations is, for
$\delta = 0.5$, $12\%$ (analytic), $5\%$ (first order), and
$0.6\%$ (second order). On the other hand, for $\delta = 0.9$, we
have $40\%$ (analytic), $16\%$ (first order), and $4\%$ (second
order).

In Fig.\ref{cuasifig2} we present results for smaller values of
$\alpha$, and in Fig.\ref{cuasifig3}, we show a zoom-plot in order
to appreciate better the differences.

In all cases considered, we can observe the expected hierarchy
between the different approximations: while the second order
approximation remains very similar to the exact result (within a
$5\%$ error) for $\alpha \geq 2$ and $\delta \leq 0.5$, the
difference between the first order approximation and the exact
result for $\alpha = 2$ and $\delta = 0.5$ is approximately
$20\%$.

Taking into account
that the different approximations are derived under the assumption $\delta\ll 1$,
the validity of the approximate results is, in general, better than expected.   Moreover,
the results of this section show  that the combination of analytic and numerical
results allow a much more efficient numerical evaluation of the Casimir energy.
In the particular case considered here (quasi concentric cylinders), from the
numerical point of view it is much more
convenient to consider sparse matrices concentrated on the diagonal, than
large matrices in which all elements are non vanishing.

\section{Concentric cylinders}
\label{concentric}

In this section we will derive an analytic result
for the vacuum energy in the concentric cylinders configuration valid
for small distances, beyond
PFA,  and we will present an improved numerical method to evaluate
the interaction energy at small distances for the particular
case of two concentric cylinders.

The exact formula for eccentric cylinders coincides, of course,
with the known result for the Casimir energy for concentric
cylinders ($\epsilon = 0$). Indeed, as $I_{n-m}(0) = \delta_{nm}$, in this
particular case the matrices $A_{np}^{{\rm TE,TM}}$ become
diagonal and the exact formula reduces to \cite{NJP, Mazzitelli-Sanchez}:
\begin{equation}
E_{12}^{\rm cc} = \frac{L} { 4\pi a^2} \int_{0}^{\infty} d\beta \
\beta\ln M^{\rm cc}(\beta), \label{cc}
\end{equation}
where
\begin{equation}
M^{\rm cc}(\beta)=\prod_n \left[1-\frac{I_n(\beta)K_n(\alpha
\beta)}{I_n(\alpha \beta)K_n(\beta)}\right]
\left[1-\frac{I'_n(\beta)K'_n(\alpha \beta)}{ I'_n(\alpha
\beta)K'_n(\beta)}\right] . \label{Mcc}
\end{equation}
The first factor corresponds to Dirichlet (TM) modes and the
second one to Neumann (TE) modes. The concentric-cylinders
configuration is interesting from a theoretical point of view,
since it can be used to test analytic and numerical methods. It
also has potential implications for the physics of nanotubes
\cite{PRA74,klim}.

The proximity limit $\alpha -1 \ll 1$ has already been analyzed
for the concentric case \cite{Mazzitelli-Sanchez}. In order to
compute the Casimir energy in this limit, it was necessary to use
the uniform expansion of the Bessel functions and to perform a
summation over all values of $n$. As expected, the result is equal
to the one obtained via the proximity approximation, namely
\begin{equation}
E_{12,PFA}^{\rm TE, cc}=E_{12,PFA}^{\rm TM, cc}=
\frac{1}{2}E_{12,PFA}^{\rm EM, cc}=
-\frac{\pi^3 L}{720 a^2} \frac{1}{(\alpha-1)^3} \, ,
\label{EPFAcc}
\end{equation}
and both TE and TM modes contribute with the same weight
to the total energy.

\subsection{Beyond proximity approximation: the next to
next to leading order}

We will now compute analytic corrections to the PFA given in
Eq.(\ref{EPFAcc}). Due to the simplicity of this configuration, we
will be able to obtain not only the next to leading order, but
also the next to next to leading contribution. In order to do
that, we need the uniform expansions of the Bessel functions. We
have
\begin{equation}
\frac{I_n(n y) K_n(n \alpha y)}{I_n(n \alpha y)
K_n(n y)} = \frac{(1 - \frac{u(t_\alpha )}{n})}{(1 -
\frac{u(t_1)}{n})} \frac{(1 + \frac{u(t_1)}{n})}{(1 +
\frac{u(t_\alpha)}{n})}e^{- 2 n\left[\eta (\alpha y)
- \eta (y)\right]},
\label{ue1}
\end{equation}
and
\begin{equation}
\frac{I_n'(n y) K_n'(n \alpha y)}{I_n'(n \alpha y)
K_n'(n y)} = \frac{(1 - \frac{v(t_\alpha )}{n})}{(1 -
\frac{v(t_1)}{n})} \frac{(1 + \frac{v(t_1)}{n})}{(1 +
\frac{v(t_\alpha)}{n})}e^{- 2 n\left[\eta (\alpha y)
- \eta (y)\right]},
\label{ue2}
\end{equation}
where
\begin{eqnarray}
\eta (y)&=& \sqrt{1 + y^2}+ \ln{\frac{y}{1 + \sqrt{1
+ y^2}}} ;\nonumber \\
u(t) &=& \frac{3 t - 5 t^3}{24}~;~ t_\alpha = \frac{1}
{\sqrt{1 +
\alpha^2 y^2}}, \nonumber \\
v(t) &=& \frac{7 t^3 - 9 t}{24}. \label{uv}
\end{eqnarray}

With these expansions at hand, we can evaluate the
matrix $M$
both for the TE and TM modes. The expression in Eq.
(\ref{Mcc}) can be
approximated by
\begin{equation}
M^{cc} \approx \left(1 - e^{-2 n \Delta\eta(y)}
A(n,y)\right)\left(1 - e^{-2 n \Delta\eta(y)} B(n,y)\right),
\label{Mcc2}\end{equation}
where
\begin{equation}
\Delta\eta(y) = h(y) (\alpha - 1) - \frac{(\alpha
- 1)^2}{2h(y)} + \frac{(2 + 3 y^2)}{6h(y)^3} (\alpha - 1)^3,
\label{deltaeta}
\end{equation}
with $h(y) = \sqrt{1 + y^2}$. In Eq.(\ref{Mcc2})
we have defined coefficients $A(n,y)$ and
$B(n,y)$ in terms of the expansions of the functions
$u(t)$ and $v(t)$. They
read
\begin{eqnarray}
A(n,y) &=& 1 + (\alpha - 1) \frac{y^2}{4n}\frac{(6 +
7 y^2 + y^4)}{(1 + y^2)^{\frac{7}{2}}},\nonumber \\
B(n,y) &=& 1 - (\alpha - 1) \frac{y^2}{4n}\frac{(-4 +
3 7 y^2)}{(1 + y^2)^{\frac{5}{2}}}.
\label{ab}
\end{eqnarray}

Replacing Eq.(\ref{Mcc2}) into Eq.(\ref{cc}), and
expanding the logarithm as a series,
it is possible to compute explicitly the remaining
integrals in $\beta$. After a long calculation,
the Casimir energy can be written  as
\begin{eqnarray}
E_{12}^{\rm cc} &\approx& -\frac{\pi^3 L}{360 a^2
(\alpha -
1)^3}\bigg\{1 +( \frac{1}{4}+\frac{1}{4} )(\alpha - 1) \nonumber \\
&- & \left(\frac{1}{\pi^2} + \frac{1}{\pi^2} +  \frac{1}{10} \right)
(\alpha - 1)^2 + ...\bigg\}
.\label{ntntlead}\end{eqnarray}
In the expression above, the first term inside the parenthesis
corresponds  to the proximity approximation contribution given in
Eq.(\ref{EPFAcc}), while the second and third terms are the first
and second order corrections, respectively. It is worth noticing that the
sub-leading term coincides with the result obtained by means of the
semiclassical approximation \cite{Mazzitelli-Sanchez}. It is also important to
remark that both TM and TE modes contribute with the same
weight to the energy in the leading and  the next to leading orders.
However, this is not the case in the quadratic term. There is a factor  $1/\pi^2$
coming from the TM mode, and a factor $1/\pi^2 + 1/10$
corresponding to the TE one.

\subsection{Improving the convergence of the numerical evaluation}

Numerical calculations of the Casimir energy for $\alpha$ very
close to one are very difficult since  big number of terms have to
be considered in the sums, and therefore convergence problems
arise, mainly produced by underflows and overflows  in the
evaluation of Bessel functions of large orders.

In order to perform a numerical evaluation of the Casimir energy
in the proximity region, we will describe a subtraction
method, in which we have used the value of the energy in the
PFA to improve the numerics  \cite{proc08}.

\begin{figure}[!ht]
\includegraphics[width=8.7cm]{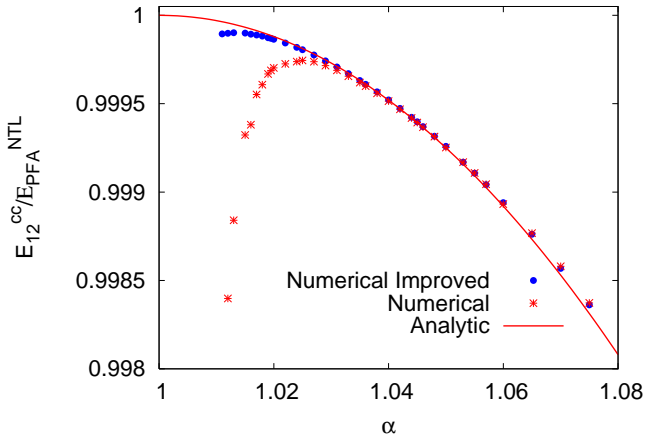}
\caption{Ratio between the exact Casimir energy for concentric cylinders
 $E_{12}^{cc}$ and the Casimir energy estimated using the PFA up to the next
to leading order $E_{PFA}^{NTL}$, as a function of
the parameter $\alpha$. This is done
for two different methods: the numerical (of slow convergence)
and the numerical improved (subtraction method).}
\label{fig6}
\end{figure}

In the case we are concerned here, we can add and subtract the
interaction energy for concentric cylinders computed using the
leading uniform asymptotic expansion for the Bessel functions, up
to first order in $\alpha-1$:
\begin{eqnarray}
\frac{K_n(n \alpha y)}{K_n(n y)} \frac{I_n(n y)}{I_n(n \alpha y)}
&\simeq& \frac{K_n'(n \alpha y)}{K_n'(n y)} \frac{I_n'(n y)}{I_n'(n \alpha y)} \nonumber\\
&\simeq &e^{-2 n(\alpha - 1)\sqrt{1+y^2}}.
\end{eqnarray}
We denote by $\tilde{E}$ the interaction energy obtained by
inserting these expansions into Eq. (\ref{cc}), which can be
computed analytically
\begin{eqnarray}
{\tilde E} &=& -\frac{1}{2(\alpha - 1)^2} \sum_{k\geq 1}
\frac{1}{k^3}\frac{1}{(e^{2k(\alpha - 1)} - 1)}
\nonumber \\
&\times & \left[1 + \frac{2 k (\alpha - 1) e^{2 k (\alpha -
1)}}{(e^{2k(\alpha - 1)}- 1)}\right],
\end{eqnarray}
and contains the leading order of the Casimir energy. Now we write
\begin{equation}
 E_{12}^{cc}=(E_{12}^{cc} - \tilde{E}) +\tilde{E}. \label{Emodificada}
\end{equation}
The difference contained in the brackets in Eq.
(\ref{Emodificada}), has a faster convergence than the original
sum and, therefore, can be easily calculated numerically.

In Fig.\ref{fig6} we present both Casimir
energy of the concentric cylinders for the direct numerical calculation (of
slow convergence) and the alternative method mentioned above. In this figure
we plot the ratio $E_{12}^{cc}/E_{PFA}^{NTL}$ where

\begin{equation}
E_{PFA}^{NTL} = -\frac{\pi^3 L}{360 a^2(\alpha -
1)^3}\left\{1 + \frac{1}{2}(\alpha - 1)\right\}.\label{ntl}
\end{equation}
As can be seen, with this subtraction method it is possible to
compute the exact energy for values of $\alpha$ much closer to
$1$, while the accuracy of the direct calculation is worse for
$\alpha < 1.02$~. Moreover, the numerical results confirm
the analytic result given in Eq.(\ref{ntntlead}). We have fit
the ratio between the Casimir interaction energy with
the next to leading correction of Eq.(\ref{ntl}),
$E_{12}^{cc}/E_{PFA}^{NTL} = a + b (\alpha - 1)^2$, obtaining $a= 1.00$, and $b = 0.29$.
The analytical results, expected from Eqs.(\ref{ntntlead}) and (\ref{ntl}) are
$a = 1$ and $b = 0.3026$, which means that we are checking the next to
next leading correction with an error smaller than $1.5\%$.

A similar method could in principle be applied to the eccentric
cylinders or the cylinder-plane configurations, although in
these cases the main difficulty is the analytic evaluation of the
approximate energy that has to be added and subtracted.

\section{A cylinder in front of a plane}
\label{cp}

\begin{figure}[!ht]
\centering
\includegraphics[width=8cm]{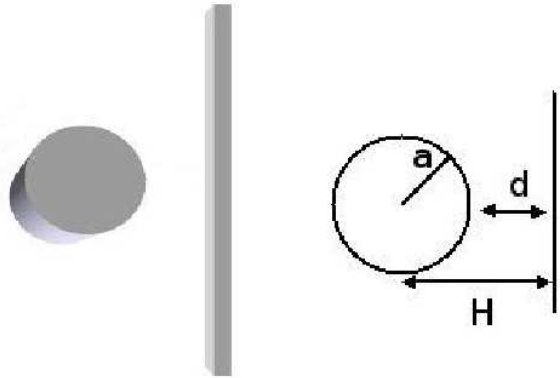}
\caption{Cylinder-plane configuration. A perfectly conducting
 cylinder of radius $a$ is in front of a perfectly conducting plane
at a distance $d$.}
\label{cpfig}
\end{figure}

In this section, we will study the cylinder-plane configuration
(Fig.\ref{cpfig}).  The Casimir energy for this configuration was first
evaluated in the PFA in Ref.\cite{europhysics}. The exact formula
has been derived in Refs.\cite{Emig, Bordag}, and has the same
structure than Eq.(\ref{E12}), but with the matrix elements  $A_{n,p}^{\rm TE}$ and
$A_{n,p}^{\rm TM}$ replaced by the corresponding ones for this geometry,
that we will denote by $A_{n,p}^{\rm TE, CP}$ and $A_{n,p}^{\rm TE, CP}$.

As suggested by simple geometric arguments, the eccentric cylinders
formula reproduces the cylinder-plane
matrix elements in the limit $b,\epsilon \rightarrow \infty$,
keeping $H=b-\epsilon$ fixed. Indeed, using the uniform expansion of
the Bessel functions, and as explained in
Ref.\cite{NJP}, the matrix elements  $A_{n,p}^{\rm TE}$ and $A_{n,p}^{\rm TM}$,
 become respectively,
\begin{eqnarray}
A_{n,p}^{\rm TE} &\simeq& - \frac{I'_n(\beta)}{K'_n(\beta)}
K_{n+p}(2 \beta H/a) \equiv A_{n,p}^{\rm TE, CP},
\label{AcpTE}
\end{eqnarray} and
\begin{eqnarray}
A_{n,p}^{\rm TM} &\simeq& \frac{I_n(\beta)}{K_n(\beta)}
K_{n+p}(2 \beta H/a) \equiv A_{n,p}^{\rm TM, CP}.
\label{AcpTM}
\end{eqnarray}
These expressions coincide with
 the result for the $\rm TE$ and $\rm TM$ modes in the cylinder
plane configuration \cite{Emig,Bordag}.

In the following we will, firstly, compare the exact Casimir
interaction energy for the cylinder-plane configuration with that
of the two eccentric cylinders,  in the limit that the latter
reproduces the former
configuration. In the end of the section, we will
numerically evaluate the cylinder plane Casimir for small distances,
in order to discuss the leading correction  to the PFA.

\subsection{Comparison between eccentric cylinders and cylinder-plane configurations}

We will now show explicitly that the numerical evaluation of
the vacuum energy for eccentric cylinders, based on
Eqs.(\ref{E12},\ref{FullTM}) and (\ref{FullTE}), reproduce the
exact results for the cylinder-plane configuration, described
by Eq.(\ref{E12}) with the matrix elements given by
Eqs.(\ref{AcpTE}) and (\ref{AcpTM}).

\begin{figure}[!ht]
\centering
\includegraphics[width=8.7cm]{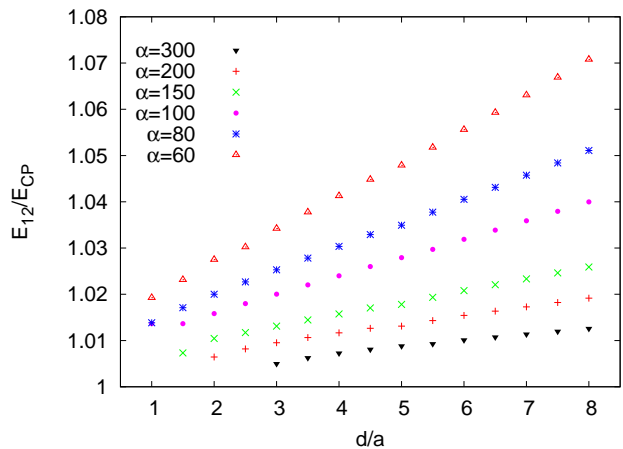}
\caption{Comparison between the exact Casimir interaction
energy for eccentric cylinders and the cylinder-plane configuration
as a function of $d=H-a$ for different values of $\alpha$.}
\label{ExandCPfig1}
\end{figure}

In Fig.\ref{ExandCPfig1} we present the ratio of both energies as
a function of $d$ for different values of $\alpha$. These runs
were done by the use of matrices of dimension $(81,81)$ and $901$
addends in the sums of Eqs.(\ref{FullTM}) and (\ref{FullTE}). The
need of  big matrices is set by the cylinder-plane configuration
program,  while the number of addends is set by the eccentric
cylinders geometry. From the numerical results we see that, as
expected, the vacuum energy of the eccentric cylinders tends to
the vacuum energy of the cylinder in front of a plane for large
values of $b$ and $\epsilon$, when $H$ and $a$ are fixed. The
coincidence is of course better for smaller values of $d$, the
minimum distance between surfaces.

In all cases, we can see that the exact Casimir interaction for eccentric cylinders is
bigger than the cylinder plane energy. This result is expected from the PFA,
since the conducting
surfaces are closer to each other in the case of the two eccentric cylinders than
in the cylinder-plane configuration, for a given minimum distance between surfaces.

\subsection{Numerical evaluation of the cylinder-plane Casimir
energy beyond the Proximity Force Approximation}

\begin{figure}[!ht]
\centering
\includegraphics[width=8.7cm]{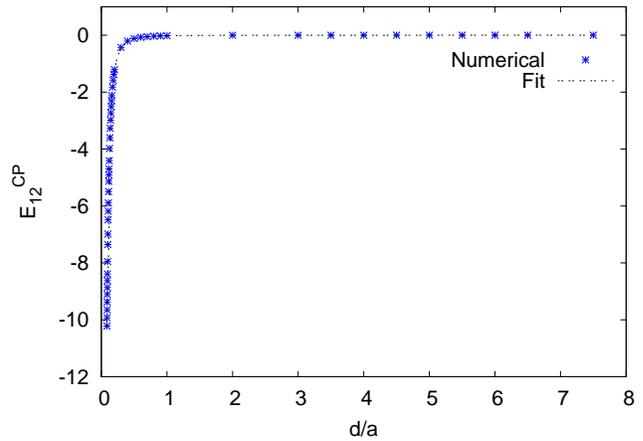}
\caption{Leading term for the Casimir interaction energy of the
cylinder-plane configuration. The leading term behaves
proportional to $-0.0228 ~d^{-5/2}$. A simple fit $f(x)=\gamma
x^{\varsigma}$ of the numerical data for $d/a<1$ gives $\gamma=
-0.021$ and $\varsigma = -2.53 $.} \label{LTcpfig1}
\end{figure}

In this section we present a detailed computation of the vacuum
energy for the cylinder-plane configuration. In Fig.\ref{LTcpfig1}
we present the Casimir interaction energy for the cylinder-plane
configuration obtained by the use of our Fortran program. For the
runs, we used a matrix of dimension (101,101) to reach the
proximity area ($d\rightarrow 0$). It must be mentioned that for
smaller values of $d$, we need to increase the dimension of the A
matrix and the integration range of $\beta$ in Eq.(\ref{E12}).
This fact becomes our major limitation to reach yet smaller values
of $d$.

We now discuss in more detail the limit $d\ll a$. This problem has
been considered from an analytical point of view in
Ref.\cite{Bordag}. Using the uniform expansions for the Bessel
functions appearing in the matrix elements $A_{n,p}^{\rm TE, CP}$
and $A_{n,p}^{\rm TE, CP}$, and after complex calculations, it can
be shown that, in the proximity limit:
\begin{equation}
E_{\rm CP}^{\rm TM}=-\frac{1}{2\pi}\sqrt{\frac{a}{d^5}}\frac{3\zeta(4)}{32\sqrt{2}}
\bigg( 1+ 0.1944 {\frac{d}{a}} + ... \bigg),
\label{E12CPBTM}
\end{equation}
\begin{equation}
E_{\rm CP}^{\rm TE}=-\frac{1}{2\pi}\sqrt{\frac{a}{d^5}}\frac{3\zeta(4)}{32\sqrt{2}}
\bigg( 1 - 1.1565 {\frac{d}{a}} + ... \bigg),
\label{E12CPBTE}
\end{equation}
where we have written separately the contributions of TM and TE modes.

\begin{figure}[!h]
\centering
\includegraphics[width=8.7cm]{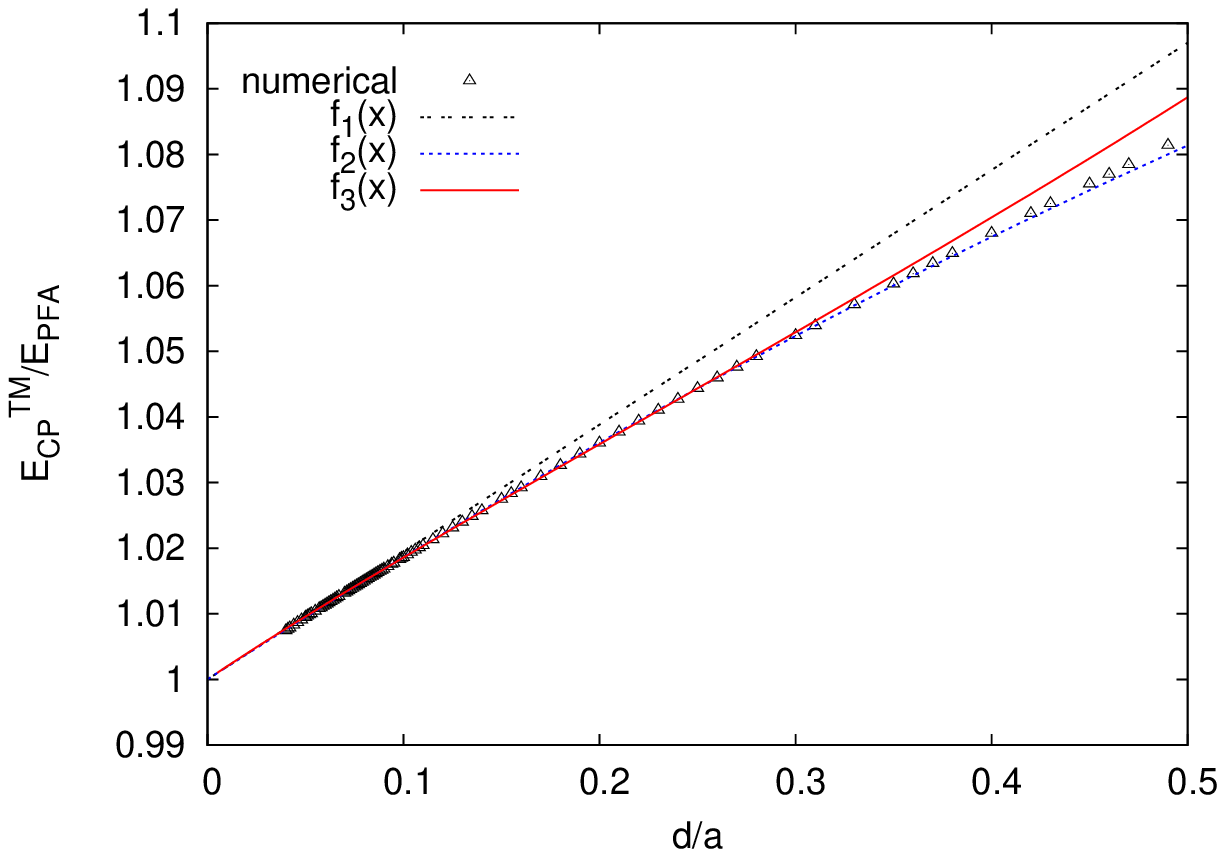}
\caption{Numerical result for the TM modes
for the cylinder-plane configuration, and the
corresponding fits presented in Table \ref{tabla1}. A simple
linear fit $f(x)= a+ b x$ of the numerical data in the
interval $0.04 \leq d/a \leq 0.07$ gives $a = 0.9999$ and
$b = 0.1900$. The theoretical values are $a = 1$ and $b = 0.1944$.}
\label{CPTM}
\end{figure}

\begin{widetext}
\begin{center}
\begin{table}
\begin{tabular}{c|c|c|c}
  $d/a$   &  $f_1(x)=1+b x$  & $f_2(x)=1+b*x+c*x^2$   & $f_3(x)= 1+b*x+c*x^2*\log(x)$   \\  \hline \hline
$[0.04:0.15]$ & $b = 0.1864$ & $b = 0.1922, c= -0.0601$ & $b= 0.1961, c=0.0438$ \\ \hline
$[0.04:0.20]$ & $b= 0.1849$ & $b= 0.1923, c= -0.0613$ & $b= 0.1983, c=0.0540$ \\  \hline
$[0.04:0.25]$ & $b= 0.1829$ & $b= 0.1922, c= -0.0601$ & $b= 0.2003, c=0.0634$ \\ \hline
$[0.04:0.30]$ & $b= 0.1811$ & $b= 0.1920, c= -0.0586$ & $b= 0.2022, c= 0.0716$ \\ \hline
$[0.04:0.35]$ & $b= 0.1794$ & $b= 0.1918, c= -0.0572$ & $b= 0.2045, c= 0.0810$ \\ \hline
$[0.04:0.40]$ & $b= 0.1771$ & $b= 0.1914, c= -0.0549$ & $b= 0.2076, c= 0.0935$ \\ \hline
 \hline
\end{tabular}\caption{Different fits for the numerical results of Fig. \ref{CPTM}.
We fix $f_i (0) = 1$ since the numerical data agree this value
with high precision.}\label{tabla1}
\end{table}

\begin{table}
\begin{tabular}{c|c|c|c}
  $d/a$   &  $f_1(x)=1+b x$  & $f_2(x)=1+b*x+c*x^2$   & $f_3(x)=1+b*x+c*x^2*\log(x)$   \\  \hline \hline
$[0.04:0.15]$ & $b = -0.8301$ & $b = -0.9704, c =  1.4499$ & $b = -1.0711 , c = -1.0852$ \\ \hline
$[0.04:0.20]$ & $b = -0.8013$ & $b = -0.9509 , c =  1.2326$ & $b = -1.0772 , c = -1.1141$ \\ \hline
$[0.04:0.25]$ & $b = -0.7683$ & $b = -0.9349 , c = 1.0794 $ & $b = -1.0890 , c = -1.1674$ \\ \hline
$[0.04:0.30]$ & $b = -0.7399$ & $b = -0.9222, c = 0.9772 $ & $b = -1.1037, c = -1.2306$ \\ \hline
$[0.04:0.35]$ & $b = -0.7158$ & $b = -0.9091, c = 0.8879$ & $b = -1.1232, c =  -1.3115$ \\ \hline
$[0.04:0.40]$ & $b = -0.6851$ & $b = -0.8943, c = 0.7999$ & $b = -1.1534, c = -1.4360$ \\\hline
 \hline
\end{tabular}\caption{Different fits for the numerical results of Fig. \ref{CPTE}.
We fix $f_i (0) = 1$ since the numerical data agree this value
with high precision.}\label{tabla2}
\end{table}\end{center}
\end{widetext}

\begin{figure}[!ht]
\centering
\includegraphics[width=8.7cm]{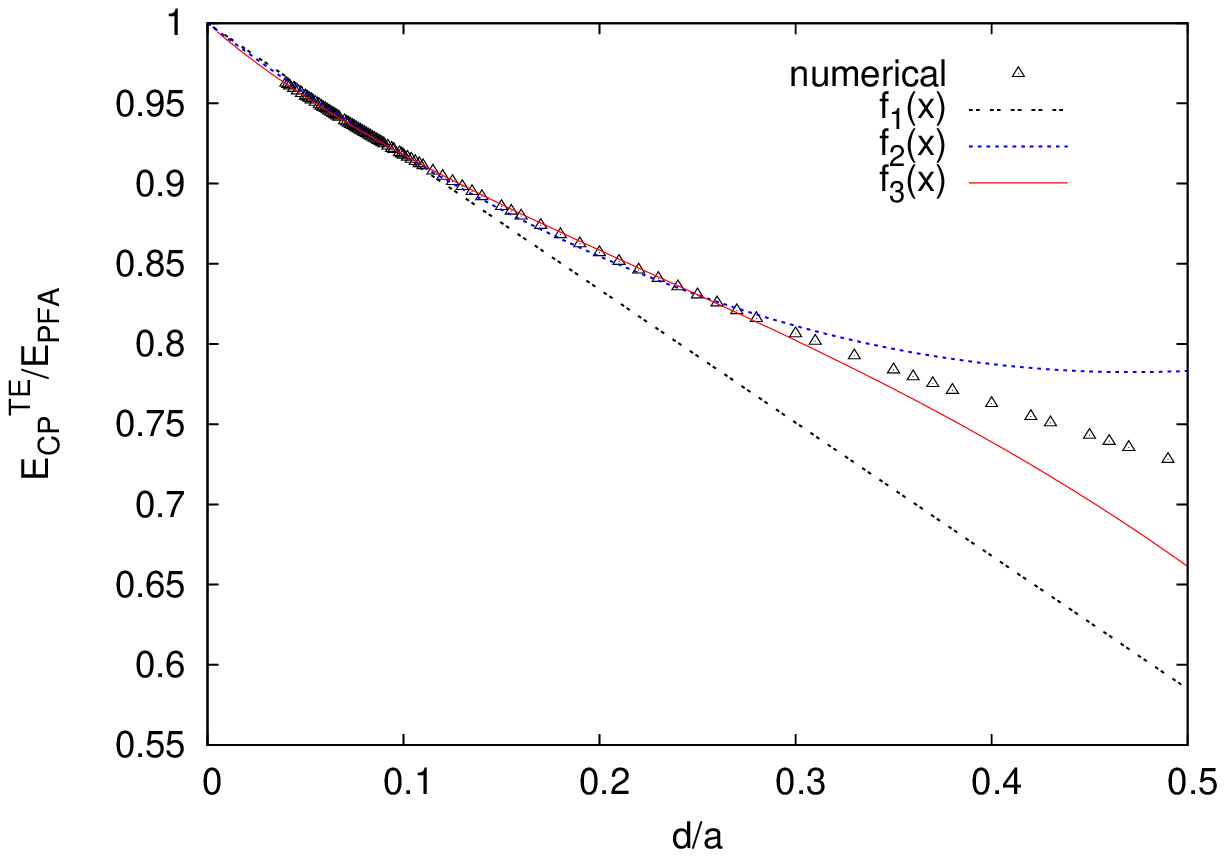}
\caption{ Numerical result for the TE modes
for the cylinder-plane configuration, and the
corresponding fits presented in Table \ref{tabla2}. A simple
linear fit $f(x)= a+ b x$ of the numerical data in the
interval $0.04 \leq d/a \leq 0.07$ gives $a = 0.9940$ and
$b = - 0.7808$. The theoretical values are $a = 1$ and $b = -1.1565$.}
\label{CPTE}
\end{figure}

\begin{figure}[!ht]
\centering
\includegraphics[width=8.7cm]{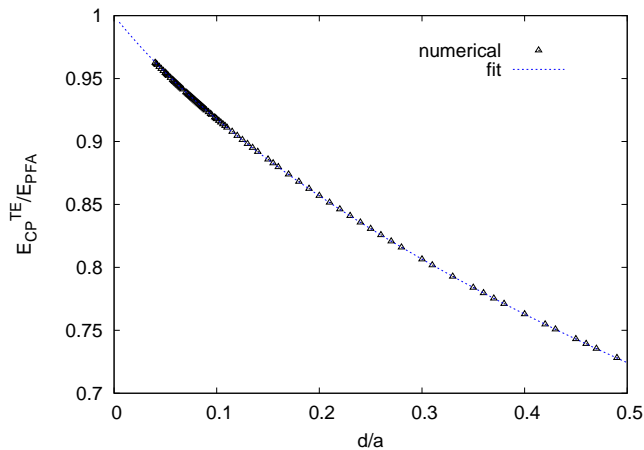}
\caption{A numerical fit of the results for the TE modes including
cubic corrections $f(x)= 1 + b x + c x^2 \log x + d x^3$. The
coefficients are $b=-1.0478$, $c= - 0.9485$, and $d = 0.6708$.}
\label{CPTM2}
\end{figure}

We will discuss the first order corrections to PFA for TM and TE
modes separately. In Fig.\ref{CPTM}, we show our numerical results
for the TM modes.  The fit of the numerical results depends of
course on the interval chosen for $d/a$. There is an obvious
compromise: on the one hand,  as already mentioned, we cannot
consider very small values for  $d/a$ because of numerical
limitations. On the other hand, the expansion in powers of $d/a$
are expected to be valid only for $d/a\ll 1$.   In any case, as
can be seen from Table I,  the different fits for the numerical
results are stable, and confirm both the PFA to leading and next
to leading orders. Indeed, the results are fully compatible with
the analytic results given in Eq.(\ref{E12CPBTM}), considering
both linear and quadratic fits of the numerical results. Moreover,
a simple linear fit in a smaller range of $d/a$ gives $a = 0.9999$
and $b= 0.1900$ and already reproduces the analytical results
\cite{Bordag} with high accuracy (see also numerical findings in
\cite{gies}).

In Fig.\ref{CPTE}, we show our results for the Neumann modes, and
we include in Table II different fits of the numerical data. In
this case, the value obtained for the linear correction to PFA
depends strongly on the assumption about the next non trivial
correction. This is not surprising: as we cannot consider
extremely small values for $d/a$, the non linear corrections may
have a non  negligible contribution in the intervals chosen for
the fits. For example, a simple linear fit gives $a = 0.994$ and
$b= -0.7808$ which does not coincide with the result in
Eq.(\ref{E12CPBTE}). However, based on the discussion about the
slower convergence of the Neumann corrections presented in
Ref.\cite{Bordag}, we have allowed the possibility of non linear
corrections proportional to $(d/a)^2 \ln(d/a)$ in our fits.
Remarkably, when this non linear corrections are taken into
account, the coefficient of the  linear correction gets closer to
the analytic prediction in Eq.(\ref{E12CPBTE}), that we reproduce
with an error less than $7\%$. Note that, as can be seen in
Fig.\ref{CPTM}, this is not the case for TM modes, since the best
fit of the numerical data contains a quadratic term without a
logarithm. In Fig.13 we show a fit of the numerical data for TM
modes that includes a cubic correction $(d/a)^3$. With this
additional term, the fit reproduces the numerical data up to
$d/a=0.5$.

To summarize, the fits of the numerical data clearly confirm the
analytic prediction for the TM modes, and suggest that the next
non trivial correction for the TE modes is not quadratic but
proportional to $(d/a)^2 \ln(d/a)$.
\section{Final remarks}

We have numerically evaluated the Casimir interaction energy for
the two eccentric cylinders configuration and for  the cylinder
plane geometry, extending in several directions the numerical
results presented in Ref.\cite{NJP}. For quasi concentric
cylinders, we have shown that the approximation based on
tridiagonal matrices derived in \cite{NJP} is in good agreement
with the numerical values. We also extended this approximation to
the case of pentadiagonal matrices. Our results show that, for
small eccentricities, it is far more efficient to consider the
contribution of the matrix elements near the diagonal, than a
"tour de force" numerical calculation based on the exact formula.

For concentric cylinders, we have obtained analytically the quadratic corrections to the PFA.
As far as we know, this is the first explicit non linear correction to PFA existing in the literature.
We have also shown that the PFA can be used as a useful tool in order to improve the
numerical evaluation at very small distances, and we have used this improvement in order
to check numerically the non linear correction to PFA.

Finally, we have analyzed in detail some numerical results for the cylinder-plane geometry. On the
one hand, we have shown that the Casimir energy for this configuration can be obtained
from that of the two eccentric cylinders. Although this coincidence has been anticipated for the matrix
elements  in Ref.\cite{NJP}, the numerical data show that the result is also valid for the energy. On the
other hand, we have computed the TE and TM contributions to the energy for small distances,
and compared the fits of the numerical results with existing analytic predictions for the linear corrections
to the PFA.

\bigskip

\acknowledgments
This work has been supported by CONICET, UBA and ANPCyT,
Argentina. We would like to thank M. Bordag for useful comments.



\begin{thebibliography}{100}
\bibitem{nota} We are using different programs to simulate the different situations
such as the eccentric cylinders, quasi concentric to first order and
quasi concentric to second order. As we have to assure convergence in
each of them, that is why we have used bigger matrix than before.
\end{thebibliography}
\end{document}